\documentstyle[aps,prd,preprint,eqsecnum]{revtex}

\newcommand{\be}{\begin{equation}}
\newcommand{\ee}{\end{equation}}
\newcommand{\ba}{\begin{eqnarray}}
\newcommand{\ea}{\end{eqnarray}}
\newcommand{\bas}{\begin{eqnarray*}}
\newcommand{\eas}{\end{eqnarray*}}
\newcommand{\hsp}{\hspace{.5cm}}

\newcommand{\spaceand}{\hsp {\rm and} \hsp}

\tightenlines

\begin{document}

\draft

\title{The Action of Instantons with Nut Charge}
\author{C.J.Hunter\thanks{email:  C.J.Hunter@damtp.cam.ac.uk}}
\address{Department of Applied Mathematics and
      Theoretical Physics, University of Cambridge,
      \\Silver Street, Cambridge CB3 9EW, United Kingdom
       }
\date{6 July 1997}

\maketitle

\begin{abstract}
We examine the effect of a non-trivial 
{\it nut charge} on the action of  
non-compact four-dimensional instantons with a $U(1)$ isometry.  
If the instanton 
action is calculated by dimensionally reducing along the isometry, then the 
nut charge is found to make an explicit non-zero contribution.  For metrics
satisfying AF, ALF or ALE boundary conditions, the action can be 
expressed entirely in terms of quantities (including the nut charge) 
defined on the fixed point set of the isometry.
A source (or sink) of nut charge also implies the 
presence of a {\it Misner string} coordinate singularity, which will have 
an important effect on the Hamiltonian of the instanton.    
\end{abstract}


\narrowtext

\section{Introduction}

Since the first publication \cite{Strominger96}
that proposed a microscopic interpretation of black hole entropy
in terms of D-branes, there has been a resurgence of interest
in black hole entropy.  From the Euclidean quantum gravity point of view
\cite{Gibbons77}, this
entropy is (in $d$-dimensions) due to the $d-2$ dimensional fixed point set of
the imaginary time isometry.  However, there are many other types of fixed
point sets which could also give rise to entropy.  

In this paper we investigate four-dimensional instantons with {\em nut charge}.
The presence of a nonzero total nut
charge means that not only are the fixed point sets more complicated than the 
simple black hole ones, but also that 
the solutions will not be asymptotically flat (AF), but rather will be
asymptotically locally flat (ALF), or asymptotically locally Euclidean (ALE).
The motivation for studying the properties of these instantons
is that it is an initial step towards understanding the effect of nut charge on
entropy (which will be given in a separate paper \cite{Hawking98}).

The outline of the paper is as follows:  we begin in section
\ref{sec:instantons} by introducing the gravitational instantons which will
serve as examples for the
calculations in the rest of the paper.
In section \ref{sec:fixed} the action formula for an instanton with a $U(1)$
isometry is dimensionally reduced to an integral over the
orbit space of the isometry.
Then in section \ref{sec:misner} the
nut charge, nut potential, and the corresponding {\em Misner string} coordinate
singularity are described.
Some concluding remarks are made in section \ref{sec:conclusions},
and a few
mathematical results used in other sections are given in the appendices.

\section{\mbox{\boldmath{$U(1)$}} Gravitational Instantons}
 \label{sec:instantons}

Much work has been devoted to the subject of black hole thermodynamics.  In
general relativity, the non-trivial value for the 
entropy of a black hole is due to the presence of the fixed point set of the
periodic imaginary time Killing vector (the fixed point set is actually 
the black hole horizon).  
Although by far the most physically relevant, 
black hole horizons are just the simplest kind of fixed point set of a
$U(1)$ isometry.  More generally, in four dimensions such fixed point sets can 
be classified into zero-dimensional nuts and two-dimensional bolts, which may
or may not have a nut charge (the black hole horizons that have been considered
so far have zero nut charge).  In this section we introduce examples of  
gravitational instantons with nuts, or with bolts with nut charge, and calculate
their actions.

When written in terms of a single coordinate system, metrics with nut charge
are found to have Misner strings, 
two-dimensional coordinate singularities, which are the result of trying to 
use stationary coordinates for metrics that are not a product near infinity.  
The presence of a Misner string means that
the metrics are not asymptotically flat, and hence 
we cannot use flat space as a background for the action and  
Hamiltonian calculations (a background metric is necessary in order to obtain
a finite result), but 
instead must consider a background with the same asymptotic behaviour.  In
our case, this will turn out to be the multi-Taub-NUT \cite{Hawking77} or the 
ALE \cite{Gibbons78} metrics.

The spaces that we will consider are the 
oriented manifolds $\cal M$, with Riemannian metrics $g_{\mu\nu}$,
that have a $U(1)$ isometry with group action $\mu_\tau :{\cal M} \rightarrow
{\cal M}$, where the group parameter is taken to be the
coordinate $\tau$ (and $\xi = \partial_\tau$ is the corresponding 
Killing vector) which has period $\beta$.  In many cases (but not all), $\tau$
will be an imaginary time coordinate obtained by analytically continuing a
Lorentzian metric to a Euclidean one.  The classification of such metrics in 
terms of their fixed point sets was first considered by
Gibbons and Hawking \cite{Gibbons79a} in 1979.

The $U(1)$ isometry has a fixed point at $p$ if $\xi=0$  at $p$.  This gives rise to an isometry
$\mu_\tau *$ of the tangent space at $p$, which is generated by the covariant
derivative of the Killing vector, i.e., the antisymmetric matrix 
$\xi_{\mu;\nu}$.  We are interested in the rank of $\xi_{\mu;\nu}$, because it 
tells us the number of directions that are invariant under $\mu_\tau*$.   
An antisymmetric $4\times 4$ matrix has rank (besides the trivial rank of zero)
two or four.  

If the rank is two, then there is a local two-dimensional surface around $p$
in which $\mu_\tau$ will be the identity (and hence the points on this surface
will be fixed).
The surface is generated by the vectors which are fixed by $\mu_\tau*$.  If
we move in the other directions, then $\mu_\tau$ is not the identity, since
$\mu_\tau*$ does not fix vectors that point in those directions.  Thus, we have
a two-dimensional fixed point set, called a {\it bolt}.  
The canonical example
is the Euclidean Schwarzschild metric, where the horizon two-surface is a bolt
of the imaginary time isometry.  

If the rank is four, then we see that there are no directions in which 
$\mu_\tau$ is the identity.  Thus we have an isolated, or zero-dimensional
fixed point set, called a {\it nut},
after its canonical example, the Euclidean Taub-NUT instanton,
which is described further in section \ref{sec:Taub}.

We would like to consider the manifold $\cal M$ in terms of a fibre bundle 
structure
based on the $U(1)$ isometry, that is, where the fibres are the orbits of $\xi$.
In order to do so, we cannot include the fixed points of $\xi$, since they
have degenerate orbits.  Thus, we want to consider ${\cal M}' = {\cal M} - 
\{{\rm fixed\;points}\}$, that is the manifold minus its fixed point set.  
Once we have deleted the degenerate fibres, we can consider $\cal M'$ 
as a $U(1)$ principal bundle over the orbit space $\Xi$.  The fact that we want
to work with ${\cal M}'$ rather than $\cal M$ will be very important in our
calculations. 

Given an instanton,
we are interested in calculating its action, since this is an 
important quantity for describing the thermodynamics of the system, and also
gives a measure of the probability for a decay mediated by the
instanton to occur.
In order to obtain a finite value
for the action of a non-compact spacetime,
it is necessary to define a background spacetime, which may be considered as a
ground state or reference spacetime, with respect to which the
spacetime under consideration may be compared \cite{Hawking96a,Hawking96b}. 
The Euclidean action is then given by\footnote{Additional terms
due to the non-orthogonality of the boundaries, described in \cite{Hawking96b},
have been omitted for clarity.}
\begin{equation}
 \label{eqn:ins_action}
 I = -\frac{1}{16\pi} \int_{\cal M} d^4x\, \sqrt{g}R -
  \frac{1}{8\pi} \int_{\partial \cal M} d^3x\,\sqrt{b}[\Theta(b) -
  \Theta(\tilde{b})],
\end{equation}
where $\cal M$ is a compact region of the spacetime, with boundary
$\partial \cal M$ (which we let tend to infinity),
$b_{\mu\nu}$ is the induced metric on $\partial \cal M$, and
$\Theta$ is the trace of the extrinsic curvature of $\partial \cal M$ in
$\cal M$.  A tilde will be used throughout to indicate a term which is
defined on the background spacetime, however, for simplicity we will generally
omit the background term from calculations.
In the limit that $\partial \cal M$ goes to infinity, a finite and
well-defined action will be obtained.  Note that for solutions of the vacuum
Einstein fields equations, such as the instantons discussed later in this 
section, the action will come entirely from the surface integral
(the background is assumed to be a solution of the field equations,
and hence never has a volume contribution).
For a background to be suitable for a given spacetime, it must be possible to
match the metric that it induces on the hypersurface $\partial \cal M$, to the
metric
induced by the spacetime on $\partial \cal M$
(at least to a sufficiently high asymptotic order, so that a
difference in the induced metrics does not affect the surface integrals).
A compact spacetime does not require the use of a background.

In order to calculate the action, we would like to simplify the integration 
by factoring out the isometry.  We can accomplish this by two possible
routes -- we can either dimensionally reduce along the orbits of the isometry, 
or we can perform a Hamiltonian decomposition.  In this paper we are only
interested in the dimensional reduction method, however the Hamiltonian
decomposition is crucial in calculating the entropy of the instanton
\cite{Hawking98}. 

In order to perform dimensional reduction, we write the metric in Kaluza-Klein
form, 
\begin{equation}
 \label{eqn:general_metric}
 ds^2 = V(d\tau + 2A_i dx^i)^2 + \frac{1}{V}\gamma_{ij}dx^idx^j, 
\end{equation}
and then the 
three-dimensional orbit space $\Xi$ is characterized by a metric 
$\gamma_{ij}$, a vector field
$A_i$, and a scalar field $V$.  The fixed points of $\xi$ will occur 
at the zeros of $V$.  
As will be
shown in section \ref{sec:fixed}, one of the principle advantages of the 
Kaluza-Klein method is that we can obtain
expressions for the action of a spacetime in terms of fixed point quantities,
such as its area, nut charge and nut potential.  

In order to dimensionally
reduce along the isometry we need a fibre bundle structure, and hence we 
need to consider $\cal M'$ rather than $\cal M$.  This will lead to
a boundary in $\Xi$ around each fixed point set, and additional contributions to
the action from the regions of the manifold that we have discarded.  
The action of a nut is zero, but
if $\cal P$ is a two-surface to be excised from the manifold $\cal M$, then 
it will have a nonzero action.
To calculate this action we
need to consider a family of tubular neighbourhoods, ${\cal T}_r$, of radius
$r$ around $\cal P$.  The action of $\cal P$ is then
obtained by taking the limit as
$r \rightarrow 0$ of the actions of the tubular neighbourhoods.  
The action of ${\cal T}_r$
is entirely due to the integral of the trace of the second fundamental form
over the hypersurface $\partial {\cal T}_r$, which is related to the radial 
derivative of the boundary volume,
\[
 I = -\frac{1}{8\pi} \lim_{r\rightarrow 0} \int_{\partial {\cal T}_r} d^3x
  \sqrt{b}\,\Theta(b) 
  = -\frac{1}{8\pi} \lim_{r \rightarrow 0}  \frac{\partial}{\partial r}
     {\rm vol}(\partial {\cal T}_r), 
\]
where $b_{\mu\nu}$ is the induced metric on $\partial {\cal T}_r$, and
$\Theta$ is its extrinsic curvature.
Using a Taylor series expansion about $\cal P$ of the hypersurface volume form,
we can calculate the boundary volume to be
\be
 {\rm vol}(\partial {\cal T}_r) = 2\pi r\,{\rm vol}({\cal P}) + r^3\int_{\cal P}
  d^2x\, F(x) + {\cal O}(r^5),
\ee
where $F(x)$ depends only on the Riemann tensor, and hence is independent of 
$r$.  Differentiating with respect to the radial distance yields
\be
 \frac{\partial}{\partial r} {\rm vol}(\partial {\cal T}_r)
 = 2\pi\,{\rm vol}({\cal P}) + 2r^2\int_{\cal P}d^2x F(x) + {\cal O}(r^4),
\ee
and hence the action is
\be
 I = -\frac{1}{8\pi} \lim_{r \rightarrow 0} 
          \frac{\partial}{\partial r} {\rm vol}(\partial {\cal T}_r)
 = -\frac{1}{4} {\rm vol}({\cal P}). 
\ee
Thus a bolt makes a contribution 
to the action of
\begin{equation}
 I = - \frac{{\cal A}}{4},
\end{equation}
where $\cal A$ is the area of the two-surface.
This contribution
must be included when ${\cal M}'$, rather than ${\cal M}$ is used to calculate 
the action.
Note that if the two-surface is non-compact, that is, 
it extends out to infinity, then it must
have a counterpart in the background spacetime, since 
the behaviour at infinity which must be matched between the two 
metrics.  In this case, there will be a contribution to the action of 
\begin{equation}
 I = - \frac{\Delta {\cal A}}{4},
\end{equation}
where $\Delta {\cal A}$ is the difference in area between the two-surface in
the spacetime and in the background.

\subsection{Asymptotic Classes}

The asymptotic behaviour of instantons that we are interested can be
divided into three classes:
\begin{description}
 \item[Asymptotically Flat (AF)] 
  The boundary at infinity is an $S^1$ bundle over an $S^2$, where the $S^1$
  fibers approach a constant length.  Such bundles are labeled by their first 
  Chern number, $c_1$.  If $c_1$ is zero, the metric is said to be 
  AF, has boundary topology $S^1 \times S^2$, and 
  approaches the metric 
  \begin{equation}
   \label{eqn:AFasympt}
   ds^2 \sim \left(1-\frac{2M}{r}\right)d\tau^2 + 
       \left(1+\frac{2M}{r}\right) d{\cal E}_3^2 + {\cal O}(r^{-2}) 
  \end{equation}
  at large radius.  $d{\cal E}_3^2$ is the flat metric on three-dimensional 
  Euclidean space, and the periodicity of $\tau$ is not fixed.  The Kerr 
  instanton is AF.
 \item[Asymptotically Locally Flat (ALF)] 
  If $c_1 = n$, then the metric is ALF, the boundary 
  topology is $S^3/{\cal Z}_n$, and it asymptotically approaches the metric 
  \begin{equation}
   \label{eqn:ALFasympt}
   ds^2 = \left(1-\frac{2M}{r}\right)(d\tau + 2aN\cos\theta d\phi)^2 + 
       \left(1+\frac{2M}{r}\right) d{\cal E}_3^2 + {\cal O}(r^{-2}), 
  \end{equation} 
  where the periodicity of $\tau$ is $8\pi aN/n$.  Both Taub-NUT and 
  Taub-Bolt are ALF metrics.
 \item[Asymptotically Locally Euclidean (ALE)] 
  A metric is ALE if the manifold has 
 a boundary $S^3/\Gamma$ at infinity,
  where $\Gamma$ is a discrete subgroup of $SO(4)$ with a free action on
  $S^3$, and the metric asymptotically approaches\footnote{In 
  \cite{Gibbons79b} an alternate asymptotic form of the ALE metric is
  given, however, since we are specifically interested in vacuum solutions,
  we can use the more restrictive metric given here.}
  \begin{equation}
   \label{eqn:ALEasympt}
   ds^2 = \left(1+\frac{M}{r^4}\right) d{\cal E}_4^2 + {\cal O}(r^{-5}),
  \end{equation} 
  where $d{\cal E}_4^2$ is the four-dimensional Euclidean metric.   The
  Eguchi-Hanson instanton is an example of an ALE metric.  
\end{description}
Asymptotically Euclidean (AE) metrics are simply ALE with 
$\Gamma = 1$.  Thus, they are really just a special case of an ALE space, and
we will not be treated as a separate class. 
   Note that from the proof of the 
   positive action conjecture \cite{Schoen79}, we know 
   that the only AE instanton is four-dimensional Euclidean space.  
We will use the term ALL spaces to refer to the combined class of 
AF, ALF and ALE metrics. 

As mentioned ealier, in order to obtain a finite result for the action,
we need to use a background metric.  Thus, it would be useful to derive the
asymptotic form of the background metrics for the asymptotic classes 
presented above.  For AF metrics,
flat space is the background.  We need to scale both the imaginary time 
and the radial coordinate in order to match flat space to 
the metric (\ref{eqn:AFasympt}) on a hypersurface of constant radius $R$.  
The matched background metric is then
\be
 \label{eqn:AFbackasympt}
 d\tilde{s}^2 = \left(1-\frac{2M}{R}\right)d\tau^2
    + \left(1+\frac{2M}{R}\right)d{\cal E}_3^2.
\ee

For ALF spaces, the background is the multi-Taub-NUT metric, which has the same
asymptotic form as (\ref{eqn:ALFasympt}), but with $\tilde{M}=\tilde{N}$. 
We want to match this background metric 
to the asymptotic metric (\ref{eqn:ALFasympt}) on a hypersurface of constant 
radius $R$.
We can do this by setting the parameter
$\tilde{N} = n(R)N$, and scaling both the imaginary time $\tilde{\tau} =
n(R)\tau$, and the radial coordinate $\tilde{r} = \lambda(R)r$,  
\be
 \label{eqn:ALFbackasympt-1}
 d\tilde{s}^2  \sim  n(R)^2\left(1-\frac{2aNn(R)}{\lambda(R) r}\right)
    (d\tau+2aN\cos\theta)^2
  + \lambda(R)^2 \left(1+\frac{2Nn(R)}{r\lambda(R)}\right)
       d{\cal E}_3^2.
\ee
If we expand $n(R)$ and $\lambda(R)$ in a Taylor series in $R^{-1}$, then we
can match the metrics to the correct order by setting
\be
 n(R) = 1 - \frac{M-N}{R} \spaceand \lambda(R) = 1 + \frac{M-N}{R}.
\ee
We can then simplify the metric (\ref{eqn:ALFbackasympt-1}) by keeping only
terms which are first order in $R^{-1}$ and $r^{-1}$,
\be
 \label{eqn:ALFbackasympt}
 d\tilde{s}^2  \sim  \left(1-\frac{2N}{r}-\frac{2(M-N)}{R} \right)
    (d\tau+2aN\cos\theta)^2
  + \left(1+\frac{2N}{r} + \frac{2(M-N)}{R} \right)
       d{\cal E}_3^2.
\ee

Finally, we want to consider the background space for an ALE instanton.  This
will simply be four-dimensional Euclidean space (no scaling is necessary),
\be
 \label{eqn:ALEasymptback}
 d\tilde{s}^2 = d{\cal E}_4^2. 
\ee
  In the remainder of this section we 
introduce the Kerr, Taub-NUT, Taub-Bolt, and Eguchi-Hanson 
instantons which we will use as examples throughout the 
paper.  

\subsection{Kerr}

A Euclidean Kerr metric can be obtained by Wick rotating both the time, and the
rotation parameter of the Lorentzian Kerr solution,
\begin{eqnarray}
 \label{eqn:kerr}
 ds^2 & = & V\left[ d\tau - 2\frac{aMr\sin^2\theta}{\Delta+a^2\sin^2\theta}
             d\phi \right]^2 \nonumber \\ 
 & &      + \frac{1}{V}\left[ \frac{\Delta + a^2\sin^2\theta}{\Delta}dr^2 +  
         (\Delta+a^2\sin^2\theta)d\theta^2 + \Delta\sin^2\theta d\phi^2
         \right],
\end{eqnarray}
where
\begin{equation}
 V = 1-\frac{2Mr}{r^2-a^2\cos^2\theta} \hspace{.5cm} {\rm and} \hspace{.5cm}
   \Delta = r^2-2Mr-a^2.
\end{equation}
The metric is asymptotically flat, since the rotation parameter, $a$, only
enters to second order in the metric.  The instanton is regular in the region
$r\geq r_+$, where
\begin{equation}
 r_+ = M+\sqrt{M^2+a^2}
\end{equation}
is the horizon of the black hole, and the bolt of the co-rotating Killing vector
$\partial_\tau + \Omega\partial_\phi$, where
\begin{equation}
 \Omega = \frac{a}{r_+^2 - a^2} = \frac{a}{2Mr_+}. 
\end{equation}
The points $(\tau,r,\theta,\phi)$ must be identified with $(\tau+2\pi\gamma, r, 
\theta, \phi + 2\pi\gamma\Omega)$, where
\begin{equation}
 \gamma = \frac{2Mr_+}{\sqrt{M^2+a^2}}.
\end{equation}
  If instead of the co-rotating vector, we
simply want to consider $\xi=\partial_\tau$, then the fixed point set is a 
nut and anti-nut (an anti-nut is a nut with negative nut charge) pair located
at $r=r_+$ and $\theta=0,\pi$, that is, at the poles of the horizon.

Because of the relation between $\tau$ and $\phi$, the coordinate ranges are 
somewhat complicated.  The total area of the 
$\tau-\phi$ plane is $4\pi\gamma$.  We can take $\phi$ to be periodic
with period $2\pi$, but then $\tau$ is not a periodic coordinate 
(except when accompanied by a 
translation in $\phi$), but simply has range $[0,2\pi\gamma]$.  
If we assume that 
$\gamma\Omega = q/p$, where $q$ and $p$ are integers, then $\tau$ can be taken
to be a periodic coordinate with period $2\pi\gamma p$, in which case $\phi$ is
no longer periodic (except when accompanied by a translation in $\tau$), but 
has a range $[0,2\pi/p]$.  Since we want $\partial_\tau$ to be a $U(1)$ 
isometry, we restrict our attention to the latter case.  

The action calculation is easy.
Since Kerr is asymptotically flat, the background is a scaled flat metric 
\begin{equation}
 \label{eqn:Kerr_back}
 ds^2 = \left(1-\frac{2M}{R}\right)d\tau^2 + dr^2 + r^2d\theta^2 + 
        r^2\sin^2\theta d\phi^2,
\end{equation}
which matches (to sufficient order) the Kerr metric on a boundary of radius $R$.
The extrinsic curvature of the Kerr metric is
\begin{equation}
 \Theta(b) = \frac{2}{R} - \frac{M}{R^2} + {\cal O}(R^{-3}),
\end{equation}
while the background value is
\begin{equation}
 \Theta(\tilde{b}) = \frac{2}{R}.
\end{equation}
Thus, the action is simply
\begin{equation}
 I = -\frac{1}{8\pi}\int d\tau d\phi \int_{0}^\pi d\theta\, R^2\sin\theta 
       \left( -\frac{M}{R^2} \right)
   = \pi \gamma M.
\end{equation}

\subsection{Taub-NUT and Taub-Bolt} 
\label{sec:Taub}

The self-dual Taub-NUT instanton \cite{Hawking77} is a non-compact 
solution of the 
field equations given by the metric 
\begin{equation}
  \label{eqn:Taub_metric}
 ds^2 = V(r)(d\tau + 2N\cos\theta d\phi)^2 + \frac{1}{V(r)}dr^2 + (r^2-N^2)
  (d\theta^2 + \sin^2\theta d\phi^2),
\end{equation}
where $V(r)$ is
\begin{equation}
 V_{TN}(r) = \frac{r-N}{r+N}.
\end{equation}
In order to make the solution regular, we consider the region 
$r \geq N$, and let the period of $\tau$ be $8\pi N$.  
It is asymptotically locally flat, since a boundary at large radius $r$
is a squashed $S^3$ -- it is an $S^1$ bundle with 
constant circumference $\beta$, (parameterized by $\tau$), over an $S^2$  
which expands with the radius $r$, (parameterized by $(\theta,\phi)$).
There is a zero of $V$ at $R=N$, and hence a fixed point set.  However, the 
$S^2$ 
parameterized by $(\theta,\phi)$ vanishes at the origin, so 
the fixed point set is zero-dimensional, that is, a nut.

The Taub-Bolt instanton \cite{Page78} is a non-self-dual, non-compact solution 
of the vacuum Euclidean Einstein equations, and is also given by the metric
(\ref{eqn:Taub_metric}).  However, 
the function $V(r)$ is different, 
\begin{equation}
 V_{TB}(r) = \frac{(r-2N)(r-N/2)}{r^2-N^2}.
\end{equation}
The solution is regular if we consider the 
region $r \geq 2N$, and let $\tau$ have period $\beta = 8\pi N$.
Asymptotically, the Taub-Bolt instanton has the same behaviour, 
i.e., ALF, as the Taub-NUT instanton.  
$V$ has a zero, and hence $\xi$ has a fixed point set, at $r=2N$.
Since the $S^2$ does not vanish there, the fixed point set is 
two-dimensional, and hence a bolt.  It has area $12\pi N^2$.
Note that it is not possible to define
fermion fields on the Taub-Bolt instanton because it
has no spin structure, however, it is possible to give it a generalized spin 
structure by coupling the fermions to an external field \cite{HawkingPope78}.

Both the Taub-NUT and Taub-Bolt instantons are obtained by Wick rotating the
Lorentzian Taub-NUT solution.  While the Lorentzian solution has two
independent parameters, $M$ and $N$, the instantons only have one, since $M$
is fixed in terms of $N$ in order to avoid a conical singularity at the origin. 
The difference in $V(r)$ between the Taub-NUT and Taub-Bolt solutions arises 
from the two different ways of fixing $M$ in terms of $N$. 

We now want to calculate the action of the Taub-Bolt instanton.
The most natural background to consider is the 
Taub-NUT solution\footnote{In reference \cite{Gauntlett97} it is suggested that
the background should be ${\cal R}^3 \times S^1$.  However, this background is
appropriate only for metrics that are asymptotically flat, and not for the more
general class of ALF metrics, because one cannot match the metrics on the 
boundary at infinity.}.
In order to calculate the action, we want to consider the compact region
$\cal M$ inside a boundary of radius $R$.  The Taub-NUT metric, as given
by equation (\ref{eqn:Taub_metric}), does not match the Taub-Bolt metric
correctly on the boundary. However, if we scale $\tilde{r}$, $\tilde{\tau}$ and
$\tilde{N}$ in terms of $R$ and $N$, then we can match
the the two metrics on the boundary.
The required transformation leaves the radial
coordinate unchanged (i.e., we use the boundary of radius $R$ in the
background spacetime), while the nut charge is given by $\tilde{N} = Nm$, where
\begin{equation}
 m = \left( 1 - \frac{N}{4R} \right).
\end{equation}
The variable $\tilde{\tau}$ is transformed by
\begin{equation}
 \tilde{\tau} \rightarrow  m\tilde{\tau},
\end{equation}
so that the new $\tilde{\tau}$ coordinate will have period $8\pi N$, identical
to the period of $\tau$ in the Taub-Bolt metric.
Thus, the correctly matched Taub-NUT metric is
\begin{equation}
 \label{eqn:Taub-NUT}
 ds^2 = m^2\tilde{V}_{TN}(d\tilde{\tau} + 
          2N\cos\tilde{\theta} d\tilde{\phi})^2 +
  \frac{1}{\tilde{V}_{TN}}dr^2 + (\tilde{r}^2 - m^2N^2)(d\tilde{\theta}^2 +
   \sin^2\tilde{\theta} d\tilde{\phi}^2),
\end{equation}
where
\begin{equation}
 \tilde{V}_{TN} = \frac{r-mN}{r+mN}.
\end{equation}

If we calculate the trace of the extrinsic curvature of the three-sphere of
radius $R$ in the Taub-Bolt instanton, we obtain (to sufficient order)
\begin{equation}
 \label{eqn:Bolt_curvature}
 \Theta(b) = \frac{2}{R} - \frac{5}{4}\frac{N}{R^2} 
       + {\cal O}(\frac{1}{R^3}).
\end{equation}
A similar calculation in the background Taub-NUT solution yields
\begin{equation}
 \label{eqn:nut_curvature}
 \Theta(\tilde{b}) = \frac{2}{R} - \frac{N}{R^2}
       + {\cal O}(\frac{1}{R^3}).
\end{equation}
Thus, the infinite terms in the action (of order $R^{-1}$) will cancel when
the two quantities are subtracted.
Calculating the action, we obtain
\begin{equation}
 I  =  - \frac{1}{8\pi} \int_{\partial \cal M}\!d^3x\,\sqrt{b}[\Theta(b) -
  \Theta(\tilde{b})] =  \pi N^2.
\end{equation}

In \cite{Eguchi80}, the actions of the Taub-Bolt and Taub-Nut instantons are
given as $5\pi N^2$, and $4\pi N^2$ respectively.  These calculations 
are incorrect because they merely discard the `flat space' term, 
$2/R$, from equations (\ref{eqn:Bolt_curvature}) and 
(\ref{eqn:nut_curvature}) in order to obtain a finite value for the action.  
Simply discarding the divergent term is equivalent to a performing 
background subtraction using a reference spacetime that 
either does not match the instantons on the
boundary, or does not satisfy the field equations, and hence is not as 
physically relevant as our method of calculating the action.

\subsection{Eguchi-Hanson}
 \label{sec:eguchi}

A non-compact instanton which is a limiting case of the Taub-NUT solution 
is the Eguchi-Hanson metric \cite{Eguchi78},
\begin{equation}
 \label{eqn:Eguchi-Hanson}
 ds^2 = (1-\frac{N^4}{r^4})(\frac{r}{8N})^2(d\tau + 4N\cos\theta d\phi)^2
 + (1-\frac{N^4}{r^4})^{-1}dr^2 + \frac{1}{4}r^2d\Omega^2,
\end{equation}
which has one free parameter $N$.
The instanton is regular if we
consider the region $r \geq N$, and we let $\tau$ 
have period $8\pi N$.   
Unlike the Taub-Bolt and Taub-NUT instantons, the Eguchi-Hanson metric is 
asymptotically locally Euclidean, because as we take $r$ to infinity, the
boundary is an $S^3$ whose radius is proportional to $r$, as opposed to
 the squashed 
$S^3$ of the Taub-Bolt and Taub-NUT solutions, 
where one direction reaches a finite size. 
In addition, since the range of $\tau/4N$ is $2\pi$ and not $4\pi$, a 
surface of constant $r$ is $S^3/{\cal Z}_2$, rather than $S^3$, that is, it 
is a three-sphere with antipodal points identified.   
$\xi$ has a fixed point at $r=N$, which is a bolt of area $\pi N^2$. 

If we want to calculate the action of the spacetime, then since it is 
non-compact we need to specify a background.  Somewhat surprisingly, it 
turns out that if we are given an Eguchi-Hanson metric with parameter $N$,
then any other Eguchi-Hanson metric with parameter $M$ will be a suitable 
background.  This implies that the action must be independent of the parameter
$N$, and since this is the only parameter in the problem, that the action 
must be zero (when the background subtraction is made).   

In order to match the two metrics on a surface of constant radius $R$, 
we must rescale the imaginary time coordinate in the background,
\begin{equation}
 \tilde{\tau} \rightarrow \frac{N}{M}\tilde{\tau},
\end{equation}
so that it has the same period, $8\pi N$, as the solution that we are matching
to.  The metric is then
\begin{equation}
 \label{eqn:Eguchi_matched}
 ds^2 = (1-\frac{M^4}{\tilde{r}^4})(\frac{\tilde{r}}{8N})^2(d\tilde{\tau} + 
4N\cos\tilde{\theta} d\tilde{\phi})^2
 + (1-\frac{M^4}{\tilde{r}^4})^{-1}d\tilde{r}^2 + \frac{1}{4}\tilde{r}^2
    d\tilde{\Omega}^2.
\end{equation}
Note that while the match between the metrics 
is not perfect on a boundary of radius $R$, it is sufficiently close so
 that any difference is unimportant as $R$ is taken to infinity.

If we calculate the trace of the second fundamental forms we obtain
\begin{equation} 
 \Theta = \frac{3R^4 - N^4}{R^3\sqrt{R^4-N^4}} 
    = \frac{3}{R} + \frac{N^4}{2R^5} + {\cal O}(\frac{1}{R^9}),
\end{equation} 
for the metric, and 
\begin{equation} 
 \tilde{\Theta} = \frac{3R^4 - M^4}{R^3\sqrt{R^4-M^4}}
    = \frac{3}{R} + \frac{M^4}{2R^5} + {\cal O}(\frac{1}{R^9}),
\end{equation} 
for the background.  When we integrate to calculate the action, the 
infinite contributions from the different metrics cancel, and since
the area element is proportional to $r^3$, the integral vanishes as the 
boundary is
taken to infinity,
\begin{equation}
 I = -\frac{1}{8\pi}\int_{\partial \cal M}  d^3x\, 
          \sqrt{b}[\Theta(b) - \Theta(\tilde{b})] = 0. 
\end{equation}

We could also use flat Euclidean space with antipodal points identified as a
background metric.  In appropriate coordinates, the metric is
\begin{equation}
 ds^2 = \frac{r^2}{(8N)^2}(d\tau + 4N\cos\theta d\phi)^2 + dr^2 + 
     \frac{1}{4}r^2d\Omega^2,
\end{equation} 
where the period of $\tau$ is $8\pi N$.  This corresponds to setting $M=0$
in the previous example.  The extrinsic curvature is
\begin{equation}
 \tilde{\Theta} = \frac{3}{R},
\end{equation}
and hence the action will still be zero.

\section{Dimensionally Reduced Action}
\label{sec:fixed}

We will now dimensionally reduce the action integral along the orbits of the 
$U(1)$ isometry.  The advantage of
this decomposition is that it allows us to express the action in terms of
geometric properties of the fixed point sets of the isometry.  For example,
we will show that for ALL metrics the action can be expressed as 
\begin{equation}
 \bar{I} =  \frac{\beta}{4} \sum_{\rm fps} \psi {\cal N} + 
        \sum_{\rm bolts} \frac{A}{4},
\end{equation}
where the subscript {\it fps} implies a sum over the fixed point set, and 
{\it bolts} implies a sum over only the bolts.  $\cal N$ and $\psi$ are the 
nut charge and nut potential, which will be defined later in this section, 
while 
$A$ is the area of the bolt.  This expression for the action 
can then be used to obtain a formula for the {\it Komar mass} of the 
instanton, 
\begin{equation}
 M_{\rm Komar} = \frac{1}{2} \sum_{\rm fps} \psi {\cal N} +
     \frac{1}{4\pi}\sum_{\rm bolts} \kappa A, 
\end{equation} 
where $\kappa$ is the surface gravity of the bolt.

We begin by calculating a general expression for the action of a
non-compact instanton with a boundary at infinity. 
The Euclidean action is
\begin{equation}
 I = -\frac{1}{16\pi} \int_{\cal M} d^4x\sqrt{g}\, R(g) -
      \frac{1}{8\pi} \int_{\partial \cal M} d^3x\sqrt{b}\, \Theta(b)
\end{equation}
(where we are implicitly assuming a background subtraction).
However, in order to dimensionally reduce, we must be integrating over the 
manifold $\cal M'$, rather than $\cal M$.  Thus, we have to introduce internal
boundaries around each connected component of the fixed point set, excise the
fixed point set from our domain of integration and then include 
the action of these excised regions separately (as shown in section 
\ref{sec:instantons}, for a bolt the action is minus one-quarter of its area,
while for a nut it is zero).  Hence, we want to write the action as
\begin{equation}
 I = -\frac{1}{16\pi} \int_{\cal M'} d^4x\sqrt{g}\, R(g) -
      \frac{1}{8\pi} \int_{\partial \cal M'} d^3x\sqrt{b}\, \Theta(b)
      - \sum_{\rm bolts} \frac{A}{4}.
\end{equation}
We now write the metric in Kaluza-Klein form,
\begin{equation}
 ds^2 = e^{2\sigma}(d\tau + 2A_i dx^i) + e^{-2\sigma}\gamma_{ij} dx^idx^j,
\end{equation}
and note that 
the volume integrand can be expressed in terms of orbit space variables as
\begin{equation}
 \sqrt{g}R(g) = \sqrt{\gamma}[ R(\gamma) + 2D^2\sigma - 2(D\sigma)^2
     - e^{4\sigma}F^2 ],
\end{equation}
where $F_{ij}$ is the field strength associated with the one-form potential
$A_i$,
\begin{equation}
 F_{ij} = \partial_i A_j - \partial_j A_i.
\end{equation}
Thus, the volume contribution to the action is
\begin{equation}
 I_{\rm vol} =  -\frac{1}{16\pi} \int_{\cal M'} d^4x\sqrt{g}\, R(g) =
                -\frac{\beta}{16\pi} \int_{\Xi} d^3x\,\sqrt{\gamma}
         [ R(\gamma) + 2D^2\sigma - 2(D\sigma)^2 - e^{4\sigma}F^2 ].
\end{equation}
We can integrate the second derivative term to obtain
\begin{equation}
 I_{\rm vol} =  -\frac{\beta}{16\pi} \int_{\Xi} d^3x\sqrt{\gamma}\,
         [ R(\gamma) - 2(D\sigma)^2 - e^{4\sigma}F^2 ]
 - \frac{\beta}{8\pi} \int_{\partial \Xi} d^2x\sqrt{c}\, 
             v^\mu D_\mu\sigma,
\end{equation}
where $v^\mu$ is the unit normal to the boundary.

The surface integral contribution to the action is
\begin{equation}
 I_{\rm surf}=-\frac{1}{8\pi}\int_{\partial\cal M'} d^3x\sqrt{b}\,\Theta(b).
\end{equation}
We can express the integrand in terms of orbit space variables, 
\begin{equation}
 \sqrt{b}\,\Theta(b) = \sqrt{c}\,\Theta(c) - \sqrt{c}\, v^\mu
    \partial_\mu\sigma,
\end{equation}
where $c_{ab}$ is the induced metric on the boundary in the orbit space.
Thus the surface term can be written as
\begin{equation}
 I_{\rm surf}=
    -\frac{\beta}{8\pi}\int_{\partial \Xi} d^2x\,\sqrt{c}\,\Theta(c)
 + \frac{\beta}{8\pi}\int_{\partial \Xi} d^2x\sqrt{c}\,
        v^\mu\partial_\mu \sigma.
\end{equation}

If we recombine the volume and surface terms (and include the action of the 
excised regions), we find that the action is given by
\be
 I  =  -\frac{\beta}{16\pi} \int_{\Xi} d^3x\sqrt{\gamma}\,
         [ R(\gamma) - 2(D\sigma)^2 - e^{4\sigma}F^2 ]
 -\frac{\beta}{8\pi}\int_{\partial \Xi} d^2x\sqrt{c}\,\Theta(c)
       - \sum_{\rm bolts} \frac{A}{4}.
\ee
Thus, we see that we have a volume contribution which is no longer trivially
zero for solutions of the field equations, a curvature contribution from the
boundary, and a contribution from the area of the bolts.  
Consider the extrinsic curvature term on the boundary of the fixed point set.  
The curvature integral on a nut is zero, while on a bolt, as shown in appendix
\ref{app:bolt},
\begin{equation}
 -\frac{\beta}{8\pi} \int_{\rm bolt} d^2x\sqrt{c}\,\Theta(c) = \frac{A}{2}.
\end{equation} 
Thus, we can eliminate the extrinsic curvature term on the fixed point set 
boundary, and write the action as
\be
 I  =  -\frac{\beta}{16\pi} \int_{\Xi} d^3x\sqrt{\gamma}\,
         [ R(\gamma) - 2(D\sigma)^2 - e^{4\sigma}F^2 ]
 -\frac{\beta}{8\pi}\int_{\infty} d^2x\sqrt{c}\,\Theta(c)
       + \sum_{\rm bolts} \frac{A}{4},
\ee
where the integral over $\infty$ means the boundary of $\Xi$ at infinity. 

We now want to dualize the two-form field strength, $F_{ij}$, and express the
action in terms of the dual field strength,
\begin{equation}
 H_i = \frac{1}{2\sqrt{\gamma}}\gamma_{ik}\epsilon^{klm} F_{lm}.
\end{equation}
Since $F^2 = 2H^2$, we can rewrite the action as 
\be
 I  =  -\frac{\beta}{16\pi} \int_{\Xi} d^3x\sqrt{\gamma}\,
         [ R(\gamma) - 2(D\sigma)^2 - 2e^{4\sigma}H^2 ]
  -\frac{\beta}{8\pi}\int_{\infty} d^2x\sqrt{c}\,\Theta(c)
      + \sum_{\rm bolts} \frac{A}{4}.
\ee
However, this action does not reproduce the field equations because it does
not account for the Bianchi identity, $\partial_{[i}F_{jk]} = 0$, which the
field strength satisfies (since it is an exterior derivative, 
$F_{ij} = \partial_{[i}A_{j]}$).
The Bianchi identity is equivalent to a vanishing divergence of $H_i$,
\begin{equation}
 D_i H^i = 0.
\end{equation}
Since this equation does not arise by varying the action, 
we must consider a different action which contains a 
Lagrange multiplier $\psi$ (called the {\it nut potential})
 which enforces this constraint, 
\begin{eqnarray}
 I' & = & -\frac{\beta}{16\pi} \int_{\Xi} d^3x\sqrt{\gamma}\,
         [ R(\gamma) - 2(D\sigma)^2 - 2e^{4\sigma}H^2 - \psi D_i H^i]
  \nonumber \\
 & & -\frac{\beta}{8\pi}\int_{\infty} d^2x\sqrt{c}\,\Theta(c)
       + \sum_{\rm bolts} \frac{A}{4}.
\end{eqnarray}
If we vary $H^i$, then we obtain the nut potential equation of motion
\begin{equation}
 \label{eqn:nut_eqn}
 4e^{4\sigma}H_i = D_i\psi.
\end{equation}
We can use this equation to eliminate $H_i$ from the action,
\begin{eqnarray}
 I' & = & -\frac{\beta}{16\pi} \int_{\Xi} d^3x\sqrt{\gamma}\,
         [ R(\gamma) - 2(D\sigma)^2 - \frac{1}{8}e^{-4\sigma}(D\psi)^2 - 
           \frac{1}{4} \psi D_i (e^{-4\sigma}D^i \psi)]
  \nonumber \\
 & & -\frac{\beta}{8\pi}\int_{\infty} d^2x\sqrt{c}\,\Theta(c)
 + \sum_{\rm bolts} \frac{A}{4}.
\end{eqnarray}
The final term in the volume integral can be decomposed into a surface and 
volume component.  This yields
\begin{eqnarray}
 I' & = & -\frac{\beta}{16\pi} \int_{\Xi} d^3x\sqrt{\gamma}\,
         [ R(\gamma) - 2(D\sigma)^2 + \frac{1}{8}e^{-4\sigma}(D\psi)^2 ] 
  \nonumber \\
 & & + \frac{\beta}{64\pi}\int_{\partial \Xi} d^2x\sqrt{c}\, 
  \psi e^{-4\sigma} v^{i} D_i \psi 
  -\frac{\beta}{8\pi}\int_{\infty} d^2x\sqrt{c}\,\Theta(c)
     + \sum_{\rm bolts} \frac{A}{4},
\end{eqnarray}
which is a three-dimensional Einstein gravity theory coupled to 
the fields $\sigma$ and $\psi$.  Note that $\psi$ is only defined up to a 
constant, but this constant will vanish when the background subtraction is
made.  If we define the field
\begin{equation}
 V = e^{2\sigma},
\end{equation}
then we can rewrite the action as   
\begin{eqnarray}
 I' & = & -\frac{\beta}{16\pi} \int_{\Xi} d^3x\sqrt{\gamma}\,
         [ R(\gamma) - \frac{1}{2V^2}(D V)^2 + 
            \frac{1}{8V^2}(D\psi)^2 ] 
  \nonumber \\
 & & +\frac{\beta}{64\pi}\int_{\partial \Xi} d^2x\sqrt{c}\,
 \frac{1}{V^2}\psi v^{i} D_i \psi  
  -\frac{\beta}{8\pi}\int_{\infty} d^2x\sqrt{c}\,\Theta(c)
  + \sum_{\rm bolts} \frac{A}{4}.
 \label{eqn:action_psi}
\end{eqnarray}

 Using the action (\ref{eqn:action_psi}) we can derive the equations of motion
for $\gamma_{ij}$, $V$ and $\psi$,
\begin{eqnarray}
 R_{ij} - \frac{1}{2}\gamma_{ij}R & = & \frac{1}{16V^2}\left[8(D_iV)(D_jV)
        - 4(DV)^2 - 2(D_i\psi)(D_j\psi) + (D\psi)^2 \right]  \\
 VD^2V - 3(DV)^2 & = & \frac{1}{4}(D\psi)^2 \label{eqn:nut_constant} \\
 D^i\left[ \frac{1}{V^2} D_i \psi \right] & = & 0.
\end{eqnarray}
The equation for $\gamma_{ij}$ implies that the volume term vanishes, while the 
equation for $\psi$ implies that there is a a conserved quantity, the {\it nut 
charge},
\begin{equation}
 \label{eqn:nut_charge}
 {\cal N} = -\frac{1}{16\pi}\int_{S^2} d^2x \sqrt{c} 
       \frac{1}{V^2} u^i \partial_i \psi,
\end{equation}
where $S^2$ is any topological two sphere, and   
$u^i$ is an outward pointing normal (i.e., at infinity $u^i=v^i$, while on the
fixed point sets, $u^i=-v^i$).  
The nut charge
is the same for all two-surfaces which are in the same cohomology class. 
We also know (by applying the field equation for $V$) that 
the nut potential must be constant on the fixed point set boundary.  Thus, if 
we eliminate the volume term and decompose the integral on the boundary of the
fixed point set into a nut potential and nut charge term, then we obtain 
\begin{equation}
 \label{eqn:action}
 I'  = \frac{\beta}{64\pi}\int_{\infty} d^2x\sqrt{c}\,
 \frac{1}{V^2}\psi v^{i} D_i \psi  
  -\frac{\beta}{8\pi}\int_{\infty} d^2x\sqrt{c}\,\Theta(c)
  + \sum_{\rm bolts} \frac{A}{4} + \frac{\beta}{4} \sum_{\rm fps} \psi {\cal N}.
\end{equation}
This is the most general form of the action, applicable to any boundary 
conditions.  If we now restrict ourselves to ALL boundary conditions,
then we can simplify the action further.  
As shown in appendix \ref{app:Integrals}, the
surface integrals at infinity will exactly cancel in the background 
subtraction.  Thus, the action becomes 
\begin{equation}
 \label{eqn:ALaction}
 \bar{I} =  \frac{\beta}{4} \sum_{\rm fps} \psi {\cal N} + 
        \sum_{\rm bolts} \frac{A}{4},
\end{equation}
so that there is a nonzero contribution from both the the area of the bolts, 
and the nut charge and potential of the nuts and bolts.  

We now want to consider what happens to the action under
various transformations.  First, we want to consider a {\em gauge
transformation},
\be
 \label{eqn:gauge}
 \tau \rightarrow \tau + 2\lambda(x^j),
\ee
which preserves the isometry, and hence the fixed point set.  The orbit space
metric and scalar field are also unaffected by this transformation, but the 
vector potential transforms as
\be
 A_i \rightarrow A_i + \lambda_{,i}.
\ee
However, when we take the exterior derivative of $A_i$ to get the field
strength $F_{ij}$, then the $\lambda$ dependent term,
$\lambda_{,[ij]}$ is clearly zero.  Thus, both the nut charge and potential are
gauge invariant.  Hence, each term in equation (\ref{eqn:action}) for the
action is separately gauge invariant.

We see that a {\em dilation} of the scalar fields, 
\be
 \psi \rightarrow  b\psi \spaceand V  \rightarrow  bV,
\ee 
leaves the action invariant, and hence gives rise to a conserved current, the
{\it dilation current},
\begin{equation}
 J_i = \frac{1}{V}D_i V - \frac{\psi}{4V^2}D_i \psi.
\end{equation}
Since $D_iJ^i=0$, the flux of the dilation current through the boundary of
$\Xi$ must vanish,
\begin{equation}
 \int_{\partial \Xi} d^2x \sqrt{c}\,v^iJ_i = 0.
\end{equation}  
Consider the flux of the dilation current through the fixed point sets
\begin{equation}
 \frac{\beta}{16\pi}\int_{\rm fps} d^2x \sqrt{c}\,v^iJ_i = \frac{\beta}{16\pi}
     \int_{\rm fps} d^2x\sqrt{c}\,\left[ 
         v^i\frac{1}{V}D_i V - \frac{\psi}{4V^2}v^iD_i \psi\right].
\end{equation}  
The first term is equal to negative one-quarter the area of the fixed point set,
while the second term is negative one-quarter the nut charge times the nut 
potential (times $\beta$),
\begin{equation}
 \frac{\beta}{16\pi} \int_{\rm fps} d^2x \sqrt{c}\,v^iJ_i = 
 -\frac{1}{4}\sum_{\rm bolts} A - \frac{\beta}{4} \sum_{\rm fps} \psi {\cal N}.
\end{equation}  
However, in the case of ALL metrics, 
this is simply minus the action of the spacetime, so
\begin{equation}
 \bar{I} = - \frac{\beta}{16\pi}\int_{\rm fps} d^2x \sqrt{c}\,v^iJ_i, 
\end{equation}  
and hence we can convert the action into an integral at infinity,
\begin{equation}
 \bar{I} = \frac{\beta}{16\pi}\int_{\infty} d^2x \sqrt{c}\,v^iJ_i = 
      \frac{\beta}{16\pi} \int_{\infty} d^2x \sqrt{c}\,\frac{1}{V}v^iD_iV,
\end{equation}  
where we have used the fact that the integral involving the nut potential 
must vanish at infinity for ALE and ALF spaces.
This expression for the action is related to the Komar mass of the spacetime,
\begin{equation}
 \bar{I} = \frac{1}{2} \beta M_{\rm Komar},
\end{equation}
where
\begin{equation}
 M_{\rm Komar} = \frac{1}{8\pi} \int_\infty d^2\Sigma_{\mu\nu}\nabla^\mu K^\nu
 = \frac{1}{8\pi} \int_{\infty} d^2x \sqrt{c}\,\frac{1}{V}v^iD_iV.
\end{equation}
This yields a generalized Smarr formula for the instanton,
\begin{equation}
 M_{\rm Komar} = \frac{1}{4\pi}\sum_{\rm bolts} \kappa A + 
     \frac{1}{2} \sum_{\rm fps} \psi \cal N,
\end{equation} 
where $\kappa = 2\pi/\beta$ is the surface gravity of the bolt.

\section{Misner Strings and the Nut Charge}
 \label{sec:misner}

  A Misner string is a coordinate singularity which can be considered as a
manifestation of a non-trivial topological twisting of the manifold.  This
twisting is parameterized by a topological term, the nut charge, which
was defined in equation (\ref{eqn:nut_charge}).  That is, there will be Misner
strings in the spacetime if and only if there is a nonzero source of nut 
charge (although the net nut charge of the spacetime may be zero).  

The existence of a Misner string was first pointed out by Misner in his paper 
on the Lorentzian Taub-NUT spacetime \cite{Misner63}.
The Lorentzian Taub-NUT solution contains a coordinate singularity running 
along one axis.  Misner noticed that he could remove the coordinate singularity
by using different
time coordinates near the north and south poles.  However, this coordinate
transformation made the time periodic.

In the orbit space $\Sigma$, the Misner string is a line such that the  
integral of $A_i$ around a closed loop which encircles the
string does not vanish as the area of loop is taken to zero -- it 
is the Dirac string for the Kaluza-Klein vector potential $A_i$.
The curvature is well-behaved along the string, implying
that it is only a coordinate singularity.  However, it is impossible to remove
the singularity with coordinates which are adapted to the isometry -- a 
gauge transformation which preserves the isometry, as given by equation
(\ref{eqn:gauge}),  simply moves the 
singularity to another location (and doesn't affect the nut charge).  
But, by considering two coordinate
charts (which have the string in different locations) it is 
possible to obtain a non-singular atlas for the manifold.

In equation (\ref{eqn:nut_charge}), the nut charge of a two-surface
is expressed in terms of the nut potential, $\psi$.
Instead, we could write the nut charge in terms of the field strength 
$F_{ij}$,  
\begin{equation}
 \label{eqn:nut_charge2}
 {\cal N} = -\frac{1}{4\pi} \int_{\cal S} {\bf F},
 \end{equation}
where $\cal S$ is any topological two-sphere in the orbit space.  Since $\bf F$
is related to the first Chern form, ${\bf c_1} = -2\beta^{-1}{\bf F}$, we can 
also write the nut charge in terms of the first Chern number, $C_1$ 
(which is an integer), 
\begin{equation}
{\cal N} = \frac{1}{8\pi}\beta C_1.
\end{equation}
 
Any surfaces ${\cal S}_1$ and ${\cal S}_2$ that can be
continuously deformed into one another (i.e., belong to the same second 
homology class)
will have the same nut charge.  This is because the fixed points of $\xi$,
which determine the internal boundaries of $\Xi$ (and hence the second homology 
classes), act as sources and sinks for the nut charge.  We can think physically
of nut charge current as
flowing into or out of the boundaries around the fixed points, and 
 along the Misner 
strings (which then must terminate on a boundary of $\Xi$ -- either at a 
fixed point set or at infinity).  The nut charge of a two-surface 
is therefore equal to the current carried out of the two-surface by the Misner
strings.  Thus, a nut-antinut pair, connected
by a Misner string, will not provide any net nut charge (as measured by a 
two-surface at infinity), since the current will be entirely
contained in the Misner string connecting the two.  Only fixed points with 
Misner strings which extend to infinity will have a net nut charge.

The nut charge, as presented here, does not need to be quantized if we let the
periodicity of the fibres vary, since it is only  
the value $C_1 = 8\pi{\cal N}\beta^{-1}$ which is quantized.
However, a
quantization condition is obtained if the periodicity of the fibres is fixed.  
This will occur if the instanton is treated as the  
spacelike part of a five-dimensional theory (with a trivial timelike direction)
and the $U(1)$ isometry is used to perform a Kaluza-Klein reduction to four 
dimensions.  The periodicity of the
fibre is related to the electric charge of the resulting Kaluza-Klein theory, 
and hence the nut charge, which can be interpreted as a corresponding 
magnetic charge, can be quantized in terms of the electric charge.
For a discussion of this interpretation with respect to the Taub-NUT and the 
Kerr-Taub-Bolt instantons, see \cite{Gross83}.
 
We now calculate the nut charge and nut potential for our example spacetimes and
verify that equation (\ref{eqn:ALaction}) gives the correct value for the 
action.

\subsection{Examples}

The Kerr instanton, although it has no net nut charge, has a more complicated 
structure than either the Taub-Bolt or Eguchi-Hanson metrics, even though they
do have a net nut chanrge.  This is because Kerr has a nut-antinut pair.  We 
can see this by considering the vector potential,
\begin{equation}
 {\bf A} = -\frac{aMr\sin^2\theta}{r^2-2Mr-a^2 + a^2\sin^2\theta},
\end{equation}
and observing that in the orbit space, the fixed point sets at $r=r_+$ and 
$\theta=0,\pi$ are 
joined by a Misner string, which is the horizon collapsed down to one
dimension.  On the Misner string, $r=r_+$, and hence 
\begin{equation}
 {\bf A} = -\frac{Mr_+}{a}.
\end{equation}
Thus, the current flowing along the Misner string is
\begin{equation}
 -\frac{1}{4\pi} \int_0^{2\pi/p} d\phi\,A_\phi = \frac{Mr_+}{2ap} = 
        \frac{1}{4\Omega p},
\end{equation}
where we recall that the $\phi$ only runs from $0$ to $2\pi/p$. 
But, this gives the nut charges of the the nut-antinut pair,
\begin{equation}
 {\cal N}_{\rm np} = \frac{1}{4\Omega p} \hspace{.5cm} {\rm and} \hspace{.5cm} 
  {\cal N}_{\rm sp} = -\frac{1}{4\Omega p},
\end{equation}
where {\it np} and {\it sp} stand for the north and south poles respectively. 
The nut potential is
\begin{equation}
 \psi = \frac{4aM\cos\theta}{r^2-a^2\cos^2\theta}.
\end{equation}
Thus, if we evaluate it on the nut and antinut, we obtain
\begin{equation}
 \psi_{np} = 4M\Omega \hspace{.5cm} {\rm and} \hspace{.5cm} 
  \psi_{sp} = -4M\Omega,
\end{equation}
so that the nut and antinut make identical contributions to the action. 
Hence, remembering that the periodicity of $\tau$ is $2\pi \gamma p$, we see 
that the action is
\begin{equation}
 I = 2\times \frac{1}{4}(2\pi\gamma p)(4M\Omega)\frac{1}{4\Omega p} 
   = \pi \gamma M,
\end{equation}
as calculated previously.  Note that since the background has no fixed point
set, there is no background subtraction. 

The Taub-NUT and Taub-Bolt metrics are both of the same form
(where $V$ is different for the two metrics),
and have field strength 
\begin{equation}
 {\bf F} = - N\sin\theta\, {\bf d\theta} \wedge {\bf d\phi}.
\end{equation}
If we integrate the field strength over any two-sphere surrounding the internal
boundary (due to the fixed point set), we obtain ${\cal N} = N$.
Note that in
the Taub-NUT case, the nut charge is due to the presence of a nut, while in
the Taub-Bolt case, it arises from a bolt.  

We can calculate $\psi$ for the Taub-Bolt and Taub-NUT cases, 
\begin{equation}
 \psi(r) = \frac{(4r-5N)N}{(r-N)(r+N)}
 \hspace{.5cm} {\rm and} \hspace{.5cm} \tilde{\psi}(r) =  \frac{4N}{r+N},
\end{equation}
where we have normalized the potentials such that they vanish at 
infinity.  If we evaluate them on the fixed point sets of the two metrics we
obtain
\begin{equation}
 \psi(2N) = 1, \hspace{.5cm} {\rm and} \hspace{.5cm}
 \tilde{\psi}(N) = 2.
\end{equation}
Thus, if we substitute the nut charge and nut potential into equation 
(\ref{eqn:ALaction}), we obtain an action of
\begin{equation}
 \bar{I} = \frac{1}{4}8\pi N^2 - \frac{1}{4}16\pi N^2 + \frac{1}{4}12\pi N^2
   = \pi N^2.
\end{equation}

For the Eguchi-Hanson metric (\ref{eqn:Eguchi-Hanson}), 
we see that the field strength is 
\begin{equation}
 {\bf F} = - 2N\sin\theta\, {\bf d\theta} \wedge {\bf d\phi}.
\end{equation}
If we integrate this over any two-sphere surrounding the internal boundary, we 
get ${\cal N} = 2N$.
This nut charge is due to the bolt at $r=N$.  
The nut potentials are 
\begin{equation}
 \psi = -\frac{1}{32}\frac{r^4+N^4}{r^2N^2}, \hspace{.5cm}
 {\rm and} \hspace{.5cm}
 \tilde{\psi} = -\frac{1}{32}\frac{r^4+M^4}{r^2N^2}.
\end{equation}
Note that although these terms diverge at infinity, the difference between them
goes to zero, as explicitly shown in appendix \ref{app:Integrals} for any ALE metric.
If we evaluate the potentials on the bolt, then we obtain
\begin{equation}
 \psi(N) = -\frac{1}{16} \hspace{.5cm} {\rm and} \hspace{.5cm} \tilde{\psi}(M) 
 = -\frac{M^2}{16N^2}.
\end{equation}
Thus, the action is
\begin{equation}
 I  = \frac{1}{4}\left[\pi N^2 - \pi M^2 + 16\pi N^2\frac{M^2-N^2}{16N^2}
         \right]  = 0.
\end{equation}

\section{Conclusions}
 \label{sec:conclusions}

We have examined some of the properties of four-dimensional 
instantons which have a nut charge, and found that they contain additional
complications that the simpler black hole metrics which have been studied up
to now do not possess.  If we dimensionally reduce the action of the instanton
along the isometry, we see that it cannot be described
purely in terms of the area of the fixed point set, but must include a 
contribution from the nut charge and nut potential of the instanton, as well as
from a surface integral at infinity.  The action is given by equation
(\ref{eqn:action})
\begin{equation}
 I'  = \frac{\beta}{64\pi}\int_{\infty} d^2x\sqrt{c}\,
 \frac{1}{V^2}\psi v^{i} D_i \psi
  -\frac{\beta}{8\pi}\int_{\infty} d^2x\sqrt{c}\,\Theta(c)
  + \sum_{\rm bolts} \frac{A}{4} + \frac{\beta}{4} \sum_{\rm fps} \psi {\cal N}.
\end{equation}
Each term in this expression was found to be invariant under the
gauge transformation $\tau \rightarrow \tau + 2\lambda(x^j)$.
In the case of ALL metrics, it was also
shown that the surface integrals vanish when a background subtraction is made,
so that the action reduces to a simple sum over fixed point set quantities
that includes both area and nut terms (equation (\ref{eqn:ALaction})),
\begin{equation}
 \bar{I} =  \frac{\beta}{4} \sum_{\rm fps} \psi {\cal N} +
        \sum_{\rm bolts} \frac{A}{4}.
\end{equation}
By considering the behaviour of the action under a dilation of the scalar
fields $V$ and $\psi$, we can rewrite the action for ALL spaces
in terms of the Komar mass,
\begin{equation}
 \bar{I} = \frac{1}{2} \beta M_{\rm Komar},
\end{equation}
which then yields a generalized Smarr formula,
\begin{equation}
 M_{\rm Komar} = \frac{1}{4\pi}\sum_{\rm bolts} \kappa A +
     \frac{1}{2} \sum_{\rm fps} \psi \cal N.
\end{equation}

As mentioned in the introduction, the effect of nut charge on the entropy of
instantons is the underlying motivation for this paper.  While we have not 
directly addressed this issue, we have proved some results about the 
action of instantons with nut charge that will prove useful in tackling the
question of entropy \cite{Hawking98}.

\section{Acknowledgements}

I would like to thank S.W. Hawking for suggesting the problem and for useful
discussion.  The Association of Commonwealth Universities and the Natural
Sciences and Engineering Research Council of Canada have provided financial
support. 
\appendix

\section{Extrinsic Curvature on a Bolt}
\label{app:bolt}

We are interested in evaluating the trace of the extrinsic curvature of a bolt 
embedded in
the three dimensional orbit space $\Xi$.  Near the bolt, we can assume that
the metric is given by
\begin{equation}
 ds^2 = V(x^i)(d\tau + 2A_idx^i)^2 + \frac{C^2(x^i)}{V(x^i)}dr^2 + 
              p_{ab}dx^adx^b,
\end{equation}
where the periodicity of $\tau$ is $\beta$, $p_{ab}dx^adx^b$ is the 
(nonvanishing) metric on a topological two-sphere (which becomes the bolt at
the zero of $V$, which is assumed to be a constant $r$ surface), 
and $C(x^i)$ is constant on the bolt. 
We are interested in deriving the
condition on $\beta$ which is necessary to avoid a conical singularity in the
$\tau-r$ plane.  To this end, we can restrict our attention to the two
dimensional metric
\begin{equation}
 ds^2 = V(x^i)d\tau^2 + \frac{C^2(x^i)}{V(x^i)}dr^2.
\end{equation}
We now want to express this as a polar coordinate system. 
If we introduce a new angular coordinate,
$\psi=2\pi\tau/\beta$, which has period $2\pi$, then the metric is
\begin{equation}
 ds^2 = \frac{\beta^2}{(2\pi)^2}V(x^i)d\psi^2 + \frac{C^2(x^i)}{V(x^i)}dr^2.
\end{equation}
We can move the origin of the coordinate system to $R=0$ by introducing the
coordinate
\begin{equation}
 R = \frac{\beta}{2\pi}\sqrt{V}.
\end{equation}
The metric is then 
\begin{equation}
 ds^2 = R^2d\psi^2 + \frac{(4\pi)^2}{\beta^2}\frac{C^2}{(V')^2}dR^2,
\end{equation}
where we have ignored terms which are not in the $\psi-r$ plane, and $V'$
indicates a derivative with respect to $r$. 
In order to avoid a conical singularity as $R \rightarrow 0$ (that is, 
as we approach the bolt), we need
\begin{equation}
 \frac{V'}{C} = \frac{4\pi}{\beta}.
 \label{eqn:bolt_relation}
\end{equation}

Now we want to use this condition on $V'$ to evaluate the trace of the 
extrinsic curvature
on the bolt.
The orbit space variables are
\begin{eqnarray}
 \gamma_{ij}dx^idx^j & = & C^2dr^2 + Vp_{ij}dx^idx^j \\
 \sqrt{\gamma} & = & CV\sqrt{p} \\
 v^r & = & -C^{-1},
\end{eqnarray}
where $v^i$ is the unit inward-pointing normal to the bolt.
If we calculate the trace of the extrinsic curvature of the bolt, we obtain
\begin{eqnarray}
 \sqrt{c}\,\Theta(c) & = & \frac{\sqrt{c}}{\sqrt{\gamma}}\frac{d}{dr}
      (v^r\sqrt{\gamma}) \nonumber \\
 & = & -\frac{1}{C}\frac{\partial}{\partial r}\left[ V\sqrt{p} \right] 
         \nonumber      \\
 & = & -\left[ \frac{V'}{C}\sqrt{p} + \frac{Vp'}{2C\sqrt{p}} \right]. 
\end{eqnarray}
But, we are evaluating this term on the bolt, where we have 
conditions on both $V$ and $V'$.  That is, $V$ vanishes on the bolt, and $V'$ is 
given by equation (\ref{eqn:bolt_relation}).  Thus, the trace of the 
extrinsic curvature is
simply 
\begin{equation}
 \sqrt{c}\,\Theta(c) = -\frac{4\pi}{\beta}\sqrt{p}.
\end{equation} 
Hence, if we integrate the trace of the extrinsic curvature over the bolt, we see that
\begin{equation}
 -\frac{\beta}{8\pi}\int_{\rm bolt}d^2x\sqrt{c}\,\Theta(c) = 
    \frac{\beta}{8\pi}\int_{\rm bolt} d^2x\frac{4\pi}{\beta}\sqrt{p}
   = \frac{1}{2} \int_{\rm bolt} d^2x\,\sqrt{p},
\end{equation}
where the final integral is simply the area of the bolt.  Thus, 
\begin{equation}
 -\frac{\beta}{8\pi}\int_{\rm bolt}d^2x\sqrt{c}\,\Theta(c) = \frac{A}{2},
\end{equation}
where $A$ is the area of the bolt.

\section{Surface Integrals at Infinity}
\label{app:Integrals}

In this appendix we are concerned with integrals in the orbit space,
over the boundary at infinity, that arise in the dimensionally reduced
equation for the action 
(\ref{eqn:action}).  There are two integrals:  a curvature integral,
\be
 \label{eqn:curv_int}
 I_c =   \int_{\infty} d^2x\sqrt{c}\,[\Theta(c) - \Theta(\tilde{c})], 
\ee
and a nut integral,
\be
 \label{eqn:nut_int}
 I_\psi = \int_\infty d^2x\sqrt{c}\,\frac{1}{V^2} ( \psi v^iD_i \psi 
              - \tilde{\psi} \tilde{v}^i \tilde{D}_i \tilde{\psi} ).
\ee
We will consider these integrals for metrics which have AF, ALF and ALE
(i.e., ALL) asymptotics, as defined in section \ref{sec:instantons}.

Beginning with the AF class, 
we write the asymptotic metric (\ref{eqn:AFasympt}) 
in Kaluza-Klein form,
\be
 ds^2  \sim  \left(1-\frac{2M}{r}\right)d\tau^2
  + \left(1-\frac{2M}{r}\right)^{-1}  d{\cal E}_3^2 + {\cal O}(r^{-2}),
\ee
and hence, up to ${\cal O}(r^{-2})$, 
the orbit space metric is simply three dimensional Euclidean space. 
Calculating the trace of the extrinsic curvature of a surface of constant radius $R$, 
we find that
\be
 \sqrt{c}\,\Theta(c) \sim 2R\sin\theta + {\cal O}(R^{-1}).
\ee
In order to calculate the curvature integral (\ref{eqn:curv_int}), we also 
need the trace of the extrinsic curvature of the boundary at infinity in the background 
metric.  From (\ref{eqn:AFbackasympt}), the matched background metric is
\be
 d\tilde{s}^2 \sim \left(1-\frac{2M}{R}\right)d\tau^2 
    + \left(1-\frac{2M}{R}\right)^{-1}d{\cal E}_3^2 + {\cal O}(R^{-2}),
\ee
and hence the orbit space metric is also
three dimensional Euclidean space.  Thus, the calculation of the trace fo the
extrinsic 
curvature of the constant radius $R$ surface will be identical to the general AF
calculation done above. So,
\be
 \sqrt{c}\, \Theta(\tilde{c}) \sim 2R\sin\theta + {\cal O}(R^{-1}),
\ee
and therefore
\be
 I_c^{\rm AF} = \int_{\infty} d^2x\sqrt{c}\,[\Theta(c) - \Theta(\tilde{c})] = 0.
\ee

We now want to calculate the nut integral (\ref{eqn:nut_int}) for 
an AF metric.
Since the asymptotic form of the vector potential is ${\cal O}(r^{-2})$,
the nut potential equation (\ref{eqn:nut_eqn}) tells us that on a surface of
constant radius $R$, 
\be
 \psi  \sim  {\cal O}(R^{-2}) \spaceand 
 \partial_r \psi  \sim  {\cal O}(R^{-3}),
\ee
where we have set the arbitrary constant equal to zero.
The asymptotic behaviour of the nut integrand is therefore
\be
 \sqrt{c}\,\frac{1}{V^2}\psi v^iD_i \psi \sim {\cal O}(R^{-3}).
\ee
The nut potential of the background is 
clearly constant (since the vector potential vanishes identically), and hence 
can be set to zero.  Thus, the nut integral will vanish as we take 
$R$ to infinity,
\be
 I^{\rm AF}_\psi = \int_\infty d^2x\sqrt{c}\,\frac{1}{V^2} ( \psi v^iD_i \psi 
              - \tilde{v}^i \tilde{D}_i \tilde{\psi} ) = 0.
\ee

If we now turn our attention to the ALF class and 
write the asymptotic metric (\ref{eqn:ALFasympt}) in Kaluza-Klein form, 
\be
 ds^2  \sim  \left(1-\frac{2M}{r}\right)[d\tau + 2aN\cos\theta d\phi]^2
  + \left(1-\frac{2M}{r}\right)^{-1}  d{\cal E}_3^2 + {\cal O}(r^{-2}),
\ee
then we see (again, up to ${\cal O}(r^{-2})$) that 
the orbit space metric is three-dimensional Euclidean space. 
Since this is the same as in the AF case, the trace of the extrinsic
curvature of a constant radius $R$ surface must be the same,
\be
 \sqrt{c} \,\Theta(c) \sim 2R\sin\theta + {\cal O}(R^{-1}).
\ee
The matched  background metric (\ref{eqn:ALFbackasympt}) 
has Kaluza-Klein form 
\ba
 d\tilde{s}^2  & \sim & \left(1-\frac{2N}{r}-\frac{2(M-N)}{R} \right)
    [d\tau+2aN\cos\theta d\phi]^2 \nonumber \\
  & & + \left(1-\frac{2N}{r} - \frac{2(M-N)}{R} \right)^{-1}
       d{\cal E}_3^2 + {\cal O}(r^{-2}),
\ea
and hence its orbit space metric is also three-dimensional Euclidean space. 
Thus, the trace of the extrinsic curvature of the boundary at 
large radius $R$ is the same as the trace of the
extrinsic curvature in the previous 
three cases.  The curvature integral will therefore vanish for ALF metrics, 
\be
 I_c^{\rm ALF} = \int_{\infty} d^2x\sqrt{c}\,[\Theta(c)- \Theta(\tilde{c})] = 0.
\ee

We now want to calculate the nut integral.  Using the
one-form potential, $A_\phi = aN\cos\theta$, we can solve the   
the nut potential equation and find that on our boundary
\be
 \psi  \sim  \frac{4aN}{R} + {\cal O}(R^{-2}) \spaceand
 \partial_r \psi  \sim  -\frac{4aN}{R^2} + {\cal O}(R^{-3}).
\ee
Hence the asymptotic behaviour of the nut integrand is
\be
 \sqrt{c}\,\frac{1}{V^2}\psi v^iD_i \psi \sim {\cal O}(R^{-1}).
\ee
Clearly, the asymptotic behaviour of the background nut potential, and hence 
the background nut integrand, will be the same, i.e. ${\cal O}(R^{-1})$,
and thus the nut integral 
(\ref{eqn:nut_int}) will be ${\cal O}(R^{-1})$ at large radius $R$, and
so will vanish at infinity, 
\be
 I^{\rm ALF}_\psi = \int_\infty d^2x\sqrt{c}\,\frac{1}{V^2} ( \psi v^iD_i \psi 
              - \tilde{v}^i \tilde{D}_i \tilde{\psi} ) = 0.
\ee

Finally, we want to tackle the ALE metrics.
We write the asymptotic form of the metric (\ref{eqn:ALEasympt}) in 
Kaluza-Klein form
\ba
 ds^2 & \sim & \left(1+\frac{M}{r^4}\right) 
     \frac{r^2}{(4aN)^2}[d\tau + 2aN\cos\theta d\phi]^2 
                 + \frac{(4aN)^2}{r^2(1+M/r^4)} \nonumber \\
 & & \times \left\{ 
        \left(1+\frac{2M}{r^4}\right) \frac{r^2}{(4aN)^2}
         \left[dr^2 + \frac{r^2}{4}(d\theta^2 + \sin^2\theta d\phi^2 \right]
  \right\} + {\cal O}(r^{-5}).
\ea
The orbit space metric is
\be
 ds^2_3  \sim   \left(1+\frac{2M}{r^4}\right) 
         \left[dr^2 + \frac{r^2}{4}(d\theta^2 + \sin^2\theta d\phi^2 )\right]
      + {\cal O}(r^{-5}).
\ee
We can then 
calculate the trace of the extrinsic curvature of the constant radius $R$ surface,
\be
 \sqrt{c}\,\Theta(c) \sim \frac{R^2}{4aN}\sin\theta + {\cal O}(R^{-2}).
\ee
To obtain the trace of the extrinsic curvature of this surface in the background metric 
(\ref{eqn:ALEasymptback}), we note that the background metric
 is simply the $M=0$ case of the general ALE metric.
Thus, up to ${\cal O}(R^{-2})$, the trace of extrinsic curvatures will be the 
same, and
hence the curvature integral will vanish, 
\be
 I_c^{\rm ALE} = \int_{\infty} d^2x\sqrt{c}\,[\Theta(c)- \Theta(\tilde{c})] = 0.
\ee

Last, but not least, we need to calculate the nut potential for an ALE metric.
Substituting the vector potential, $A_\phi = aN\cos\theta$, into the nut
potential equation yields
\be
 \psi \sim -\frac{2}{(4aN)^2}\left( r^2 -\frac{M}{r^2}\right) + 
              {\cal O}(r^{-3}).
\ee
From this, we can evaluate the integrand of the nut integral over the boundary,
\be
 \label{eqn:ALEnut_int}
 \sqrt{c}\,\frac{1}{V^2}\psi v^iD_i \psi \sim \frac{2}{4aN}R^2\sin\theta +
       {\cal O}(R^{-2}).
\ee
Just as before, we can obtain the background value by simply setting $M=0$.
This yields a background nut integrand which is identical 
(up to ${\cal O}(R^{-2})$) with equation (\ref{eqn:ALEnut_int}), and 
hence the nut integral will vanish as we take the constant radius surface to
infinity,
\be
 I^{\rm ALE}_\psi = \int_\infty d^2x\sqrt{c}\,\frac{1}{V^2} ( \psi v^iD_i \psi 
              - \tilde{v}^i \tilde{D}_i \tilde{\psi} ) = 0.
\ee

In summary, we have shown that that the curvature
and nut integrals given by equations (\ref{eqn:curv_int}) and 
(\ref{eqn:nut_int}) vanish for ALL metrics.  This justifies the 
simplification of the action formula from equation (\ref{eqn:action}) to 
(\ref{eqn:ALaction}) for this class of metrics.

\end{document}